# OSNR limitations of chip-based optical frequency comb sources for WDM coherent communications

Pablo Marin-Palomo,[1] Juned N. Kemal,[1] Wolfgang Freude,[1] Sebastian Randel[1,*] and Christian Koos[1,2,**]

[1]*Institute of Photonics and Quantum Electronics (IPQ), Karlsruhe Institute of Technology (KIT), 76131 Karlsruhe, Germany*
[2]*Institute of Microstructure Technology (IMT), Karlsruhe Institute of Technology (KIT), 76344 Eggenstein-Leopoldshafen, Germany*
*\*Sebastian.randel@kit.edu, \*\*christian.koos@kit.edu*

**Optical frequency combs have the potential to become key building blocks of optical communication subsystems. The strictly equidistant, narrow-band spectral lines of a frequency comb can serve both as carriers for massively parallel data transmission and as local oscillator for coherent reception. Recent experiments have demonstrated the viability of various chip-based comb generator concepts for communication applications, offering transmission capacities of tens of Tbit/s. Here, we investigate the influence of the comb line power and of the carrier-to-noise power ratio on the performance of a frequency comb in a WDM system. We distinguish two regimes of operation depending on whether the comb source or the transmission link limits the performance of the system, i.e., defines the link reach, restricts the choice of modulation format and sets the maximum symbol rate. Finally, we investigate the achievable OSNR and channel capacity when using the tones of a soliton Kerr frequency comb as multi-wavelength carriers for WDM systems.**

## 1. Introduction

Wavelength-division multiplexing (WDM) has been used for long-haul optical fiber links over distances of hundreds or even thousands of kilometers since the 1990s [1]. For such links, the fiber infrastructure represents the most expensive asset while the costs of transceiver components are comparatively low. However, with the explosive growth of data rates across all network levels, WDM is now becoming increasingly important also for shorter links, which are deployed in much larger quantities and which are much more sensitive to cost and size of the transceiver assemblies. This evolution is, e.g., witnessed by a strong market growth in the field of so-called data center interconnects, which link two or more datacenters or parts thereof across a metropolitan-area with typical distances of less than 100 km. In the future, even shorter WDM links are likely to emerge, e.g., in the context of high-throughput campus-area networks, which connect datacenter buildings over distances of up to a few kilometers. At present, these networks still rely on parallel transmission using spatially separated channels in thousands of single-mode fibers, each of which is operated at a comparatively low data rate of at most a few hundred Gbit/s. It is foreseeable that the concept of plain spatial parallelization will soon reach its scalability limits and that it must be complemented by spectral parallelization of data streams in each fiber to sustain further increases in data rates. This may un-lock an entirely new application space for WDM techniques, in which utmost scalability of channel counts and data rates is of the highest importance.

In this context, optical frequency comb generators may play a key role as compact and robust multi-wavelength light sources that can provide large numbers of well-defined optical carriers [2-8]. A particularly important advantage of frequency combs is the fact that comb lines are inherently equidistant in frequency, hence relaxing the requirements for inter-channel guard bands and avoiding frequency control of individual lines as needed in conventional schemes that combine arrays of independent distributed-feedback (DFB) lasers. Note that these advantages do not apply only to the WDM transmitter, but also to the receiver, where an array of discrete local oscillators (LO) may be replaced by a single comb generator [6,10]. Using an LO comb further facilitates joint digital signal processing of the WDM channels, which may reduce receiver complexity and increase phase noise tolerance [11,12]. Moreover, parallel coherent reception using an LO comb with phase-locked tones might even allow to reconstruct the time-domain waveform of the overall WDM signal and thus permit compensation of impairments caused by optical nonlinearities of the transmission fiber [13].

Besides these conceptual advantages of comb-based transmission, compact footprint and cost-efficient mass production is key for future WDM transceivers. Among the various comb generator concepts, chip-scale devices are therefore of particular interest [5-10,14-19]. When combined with highly scalable photonic integrated circuits for modulation, multiplexing, routing, and reception of data signals, such devices may become the key to compact and energy-efficient WDM transceivers that can be fabricated in large quantities at low cost and that can offer transmission capacities of tens of Tbit/s per fiber.

Chip-based frequency comb, however may present certain limitations in comparison to an array of independent lasers, namely lower optical power per line and lower optical carrier-to-noise power ratio (OCNR) per line. For large line counts, optical amplifiers are needed to boost the optical power prior to modulation, which will have direct impact on the achievable optical signal-to-noise power ratio (OSNR) of the WDM transmitter [11]. Therefore, optical frequency combs introduce new limitations in comparison with using individual DFB lasers. Here we describe how the power per line and OCNR per line affect the performance of the WDM system. We distinguish two regimes of operation de-pending on whether the comb source or the transmission link limits the performance of the system, i.e., limits the link reach, restricts the choice of modulation format and sets the maximum symbol rate. Among the various chip-based comb generators for WDM applications, so-called Kerr soliton frequency combs, or bright solitons, stand out as a particularly promising option due to their large number of low-noise optical carriers and smooth spectral envelope. We thus, carry out an analysis of the achievable OSNR and channel capacity using such bright solitons for WDM applications.

## 2. OSNR limitations in a comb-based WDM system

The performance of a WDM link is often quantified by the OSNR measured at the receiver input. The OSNR is defined as the ratio of the signal power to the amplified spontaneous emission (ASE) noise power within a reference bandwidth $B_{\text{ref}} = 12.5\,\text{GHz}$, corresponding to a reference wavelength span of 0.1 nm at a center wavelength of $1.55\,\mu\text{m}$. The OSNR can be translated into the signal-to-noise ratio SNR that refers to the actual bandwidth $B_{\text{sig}}$ of a specific signal [20]. For a coherent system limited by ASE noise,

$$\text{SNR} = \frac{2B_{\text{ref}}}{pB_{\text{sig}}}\text{OSNR}. \quad (1)$$

where $p = 1$ for a single polarization signal and $p = 2$ for a polarization multiplexed signal. In the following, we will consider dual polarization signals.

In the case of using an optical frequency comb generator (FCG) as a light source, the OSNR will depend on the comb line power $P_\ell$ and the optical carrier-to-noise ratio ($\text{OCNR}_\ell$) of the $\ell$-th comb line. In analogy to the OSNR, the OCNR relates the comb line power to the power of the background noise, measured again within a reference bandwidth $B_{\text{ref}}$ centered at the comb line frequency. The noise power at the output of the FCG is given by

$$P_{\text{n},\ell} = S_{\text{n},\ell}B_{\text{ref}} = \frac{P_\ell}{\text{OCNR}_\ell}. \quad (2)$$

Here, $S_{\text{n},\ell}$ is the noise power density copolarized with the carrier. Note that we disregard the impact of phase noise in our analysis.
The comb line power and the OCNR impact the OSNR of a WDM channel and can thus limit the achievable reach and spectral efficiency. For a quantitative analysis, consider the WDM link depicted in Fig. 1 consisting of a WDM transmitter, a link with $M$ fiber spans, and a WDM receiver [11]. Note that the comb line powers $P_\ell$ of most chip-scale comb generators are usually much weaker than the power levels emitted by state-of-the-art continuous-wave laser diodes as used in conventional WDM systems. It is therefore necessary to amplify the comb lines prior to modulation by sending them through a dedicated optical amplifier ('Comb amp.') with gain $G_0 > 1$. Data are encoded onto the various carriers by a WDM modulator unit (WDM mod.), comprising a WDM demultiplexer (DEMUX), an array of dual-polarization in-phase/quadrature (I/Q) modulators, and a WDM multiplexer (MUX). The gain of the WDM modulator unit is denoted as $g_0 < 1$, and accounts for the insertion losses of the multiplexer and the demultiplexer as well as for the insertion and modulation losses of the I/Q modulators,

$$\begin{aligned}P_{\text{s},\ell}^{(0)} &= g_0 G_0 P_\ell \\ P_{\text{n},\ell}^{(0)} &= g_0[\underbrace{S_{\text{n}}G_0 B_{\text{ref}}}_{\substack{\text{amplified}\\\text{OFC noise}}} + \underbrace{F_0 hf(G_0 - 1)B_{\text{ref}}}_{\substack{\text{ASE noise of}\\\text{comb amplifier}}}].\end{aligned} \quad (3)$$

In these relations, the quantity $F_0$ is the noise figure of the amplifier after the comb source, and $hf$ is the photon energy [21].

The WDM polarization multiplexed signal is then boosted by an optical amplifier 'Post-amp.' with gain $G_1$, and launched into the first fiber section, which attenuates the power by $g_1 < 1$,

$$\begin{aligned}P_{\text{s},\ell}^{(1)} &= g_1 G_1 P_{\text{s},\ell}^{(0)} \\ P_{\text{n},\ell}^{(1)} &= g_1[G_1 P_{\text{n},\ell}^{(0)} + F_1 hf(G_1 - 1)B_{ref}]\end{aligned}. \quad (4)$$

This fiber span might be followed by ($M - 1$) additional fiber sections, each attenuating the power by $g_m < 1$ and by the same number of additional amplifiers with a gain $G_m$ such that $G_m g_m = 1$. Assuming that all $M - 1$ links are identical ($G_m = G$, $g_m = g$, $Gg = 1$, $F_m = F$ for $m = 2...M$), the signal power and the noise power after the transmission link are given by

$$\begin{aligned}P_{\text{s},\ell}^{(M)} &= P_{\text{s}} \\ P_{\text{n},\ell}^{(M)} &= P_{\text{n},\ell}^{(1)} + (M - 1)Fhf(G - 1)B_{\text{ref}}\end{aligned}. \quad (5)$$

The data from each WDM channel are recovered at the WDM receiver, which contains an optical preamplifier ('Pre-amp') with gain $G_{\text{Rx}}$ and noise figure $F_{\text{Rx}}$ and a WDM demodulation unit ('WDM demod.'). The WDM demodulation unit uses a multitude of LO tones derived from a second FCG to recover the transmitted data with an array of IQ detectors. Note that the additive noise background of the LO comb tones can be neglected with respect to that of the received signal. The LO comb is hence not considered further in the subsequent analysis. The signal power $P_{\text{s,Rx}}$ and the noise power $P_{\text{n,Rx}}$ in the reference bandwidth $B_{\text{ref}}$ at the output of the pre-amplifier are given by

$$\begin{aligned}P_{\text{s},\ell}^{(\text{Rx})} &= G_{\text{Rx}} P_{\text{s},\ell}^{(M)} \\ P_{\text{n},\ell}^{(\text{Rx})} &= G_{\text{Rx}} P_{\text{n},\ell}^{(M)} + F_{\text{Rx}} hf(G_{\text{Rx}} - 1)B_{\text{ref}}\end{aligned}. \quad (6)$$

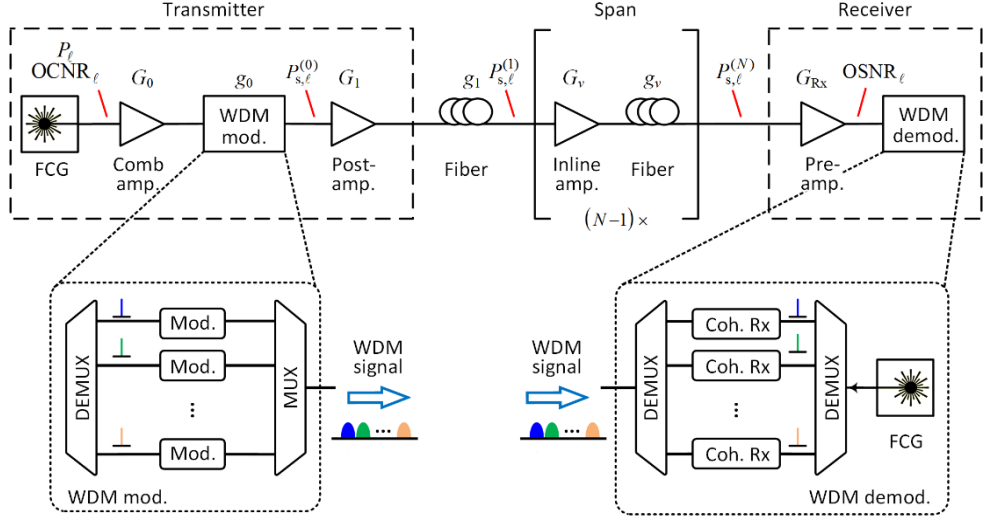

**Fig. 1**. Schematic of a WDM transmission and reception system. The comb lines of a frequency comb generator (FCG) are first amplified by an optical amplifier ('Comb amp.') before data are encoded on each line in the WDM modulation unit ('WDM mod.'), comprising a WDM demultiplexer ('DEMUX'), an array of dual-polarization I/Q modulators ('Mod.'), and a WDM multiplexer ('MUX'). The WDM signal is then boosted by an optical amplifier ('Post-amp.') and transmitted through the fiber link consisting of at least one span. Each of the $M-1$ additional spans contain an in-line amplifier ('In-line amp.') which compensates the loss of the corresponding fiber section. The data from each WDM channel are recovered at the WDM receiver, which contains a preamplifier ('Pre-amp.') and a WDM demodulation unit ('WDM demod.'). The 'WDM demod.' uses a multitude of LO tones derived from a second FCG.

To obtain the OSNR for a single WDM channel at the receiver, the ratio of the received signal power and the noise power in the reference bandwidth $B_{\text{ref}}$ needs to be calculated,

$$\text{OSNR}_\ell = \frac{P_{\text{s},\ell}^{(\text{Rx})}}{P_{\text{n},\ell}^{(\text{Rx})}}. \quad (7)$$

Introducing Eqs. (3)-(6) into Eq. (7), for a given line power $P_\ell$ and $\text{OCNR}_\ell$, the OSNR per channel can be expressed as

$$\text{OSNR}_\ell = g_1 G_1 g_0 G_0 P_\ell \left[ \underbrace{\frac{g_1 G_1 g_0 G_0 P_\ell}{\text{OCNR}_\ell}}_{\text{ampl. FCG noise}} + \underbrace{hfB_{\text{ref}} \left( \begin{array}{c} g_1 G_1 g_0 (G_0 - 1) F_0 + g_1 (G_1 - 1) F_1 \\ + g_1 (M-1)(G-1) F + \dfrac{G_{\text{Rx}} - 1}{G_{\text{Rx}}} F_{\text{Rx}} \end{array} \right)}_{\text{amplifier noise}} \right]^{-1} \quad (8)$$

In typical 32 GBd WDM links, an optical power of approximately 0 dBm per channel is launched into each fiber span for keeping the nonlinear impairments low while maintaining high OSNR [22,23]. Thus, the signal power of a single WDM channel is assumed to be 0 dBm at the beginning of each fiber span. For simplicity, the gain and the noise figure of the receiver preamplifier 'Pre-amp.' is assumed to be identical to that of the in-line amplifiers, leading to a signal power of 0 dBm per wavelength channel at the input of the receiver DEMUX. Note that for a high number of spans, the noise of the receiver preamplifier does not play a role any more, since the noise background is dominated by the contributions from the various amplifiers along the link.

Eq. (8) is the foundation of Fig. 2, which shows the OSNR as a function of the line power $P_\ell$ and the number $M$ of fiber spans for a realistic comb-based WDM system. For Fig. 2(a), the $\text{OCNR}_\ell$ of the comb lines is assumed to be infinite, i.e., the FCG itself does not introduce any practically relevant additive noise background. This can, e.g., be accomplished by dissipative Kerr soliton comb generators, see Fig. 3. The various parameters used for this study are specified in Table 1. In our analysis we consider a frequency comb source where all tones are equal in power. In Section 3, we analyze the impact of a frequency dependent power per line. For the WDM modulator unit, an overall insertion loss of 25 dB is assumed, comprising 3.5 dB of insertion loss each for the WDM demultiplexer and the multiplexer [24], 13 dB of loss for the dual-polarization IQ modulators [25], and an additional 5 dB of modulation loss, which are assumed to be independent of the modulation format for simplicity. Note that the results shown in Fig. 2(a) do not change significantly when varying these losses by a few dB.

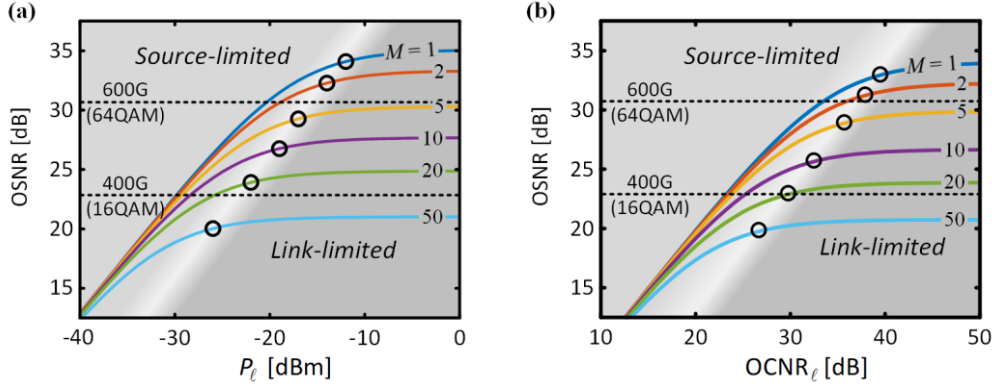

**Fig. 2.** Influence of the comb line power $P_\ell$ and of its carrier-to-noise power ratio $OCNR_\ell$ on the achievable optical signal-to-noise power ratio OSNR at the receiver for different numbers $M$ of 75 km-long spans with 0.2 dB/km loss, see Table 1 for the model parameters used. **(a)** OSNR as a function of $P_\ell$ in the limit of high $OCNR_\ell$. The plot reveals two regimes: For source-limited transmission at low line powers, the OSNR is dominated by the contribution of the comb amplifier, whereas the contributions of the various amplifiers along the link dominates for link-limited transmission at high line powers. The transition between both regimes is indicated by black circles which mark the points where the OSNR has decreased by 1 dB in comparison to its limit at high comb line powers $P_\ell$. The dashed horizontal lines indicate the minimum OSNR required for transmission of a net data rate of 400 Gbit/s using 16QAM and of 600 Gbit/s using 64QAM as a modulation format. **(b)** OSNR as a function of $OCNR_\ell$ for comb line powers $P_\ell$ that correspond to the transition points marked by circles in (a). For low $OCNR_\ell$, OSNR is dominated by the noise of the comb source, and the link performance is source-limited. For high $OCNR_\ell$, the OSNR is independent of the $OCNR_\ell$, and the transmission performance is link-limited.

The fiber spans are assumed to feature a power loss of 15 dB each, corresponding to 75 km of single-mode fiber with a propagation loss of 0.2 dB/km. The power loss of each span is exactly compensated by the 15 dB gain of the corresponding post-amplifier or in-line amplifier such that a signal launch power of 0 dBm is maintained for all spans. The gain $G_1$ of the 'Post amp.' is assumed to be the same as the gain $G_m$ of the "In-line amp.", and the gain $G_0$ of the 'Comb amp.' is adjusted such that the 0 dBm launch power is reached after the 'Post-amp.'

For low line powers $P_\ell$, Fig. 2(b) shows that the noise level at the receiver is dominated by the contribution of the comb amplifier. As a consequence, the OSNR increases in proportion to $P_\ell$ and is essentially independent of the span count $M$. In this regime, transmission performance is limited by the comb source and its associated amplifier ("source-limited"). For high comb line powers $P_\ell$, in contrast, the noise level at the receiver is dominated by the contributions of the various amplifiers along the link ("link-limited"). In this regime, the OSNR is essentially independent of $P_\ell$ and decreases with each additional span. The transition between both regimes is indicated by black circles in Fig. 2(a), indicating the points where the OSNR has decreased by 1 dB in comparison to its limit at high line power $P_\ell$. When using frequency combs in WDM systems, operation in the link-limited regime is preferred. Depending on the number of spans, this requires a minimum comb line power between –25 dBm and –

**Table 1.** Model parameters used for generating Fig. 2(a) and 2(b) from Eq. (8). The fiber attenuation , amplifier gain , and noise figures are assumed to be the same for all M fiber links ($m = 1...M$).

| Variable | Description | Value | |
|---|---|---|---|
| $B_{ref}$ | Reference bandwidth for OSNR and OCNR calculation | 12.5 GHz | |
| $P_s$ | Signal power per WDM channel at the input of each fiber span | 1 mW | (0 dBm) |
| $g_0$ | Power transmission factor of WDM modulator unit | $3.2 \times 10^{-3}$ | (−25 dB) |
| $G_0$ | Power gain factor of comb amplifier | $P_s / (P_\ell g_0 G)$ | |
| $F_0$ | Noise figure of comb amplifier | 3.2 | (5 dB) |
| $g_m = g$ | Power transmission factor of fiber section | $3.2 \times 10^{-2}$ | (−15 dB) |
| $G_m = G = g^{-1}$ | Power gain factor of post- and inline-amplifier | 32 | (15 dB) |
| $F_m = F$ | Noise figure of post- and inline-amplifiers | 3.2 | (5 dB) |
| $G_{Rx}$ | Power gain factor of pre-amplifier | 32 | (15 dB) |
| $F_{Rx}$ | Noise figure of pre-amplifier | 3.2 | (5 dB) |

15 dBm. Note that, in principle, the gain $G_1$ of the post amplifier could be decreased by increasing the gain $G_0$ of the comb amplifier. For source-limited transmission, the overall impact would be small since the OSNR is mainly dictated by the comparatively low line power $P_\ell$ that enters the comb amplifier. For link-limited transmission over a small number of spans, the OSNR can be slightly improved by decreasing the gain $G_1$ and increasing the gain $G_0$ while maintaining power levels of about 10 dBm per line at the output of the comb amplifier.

Figure 2(a) can be used as a guide for estimating the performance requirements for comb sources in WDM applications. As a reference, the plot indicates the minimum OSNR required for transmission of a net data rate of 400 Gbit/s and 600 Gbit/s per WDM channel, using 16QAM and 64QAM as a modulation format, respectively. In both cases, advanced forward-error correction schemes with 11 % overhead and BER thresholds of $1.2 \times 10^{-2}$ [26] are assumed, requiring a symbol rate of 56 GBd to provide the specified net data rates.

For many practically relevant comb sources, the $OCNR_\ell$ is finite, which will further decrease the transmission performance. Figure 2(b) shows the OSNR as a function of $OCNR_\ell$ for various span counts $M$. In this plot, the comb line power $P_\ell$ was assumed to correspond to the minimum value required for link-limited transmission, as indicated by the corresponding transition points in Fig. 2(a). For low $OCNR_\ell$, the noise level at the receiver is dominated by the noise background of the comb source, and the $OCNR_\ell$ and OSNR are essentially identical. In this case, the link performance is again source-limited. For high $OCNR_\ell$, the OSNR is independent of the $OCNR_\ell$, and the transmission performance is link-limited. Depending on the number of spans, the minimum $OCNR_\ell$ values needed for link-limited transmission range between 25 dB and 40 dB. The transition between both regimes is again marked by black circles in Fig. 2(b), indicating the points where the OSNR has decreased by 1 dB in comparison to its respective limit at high $OCNR_\ell$.

## 3. Soliton Kerr combs for massively parallel WDM communications

Chip-scale comb sources have been used in a variety of WDM transmission experiments, relying on different comb generation approaches and covering a wide range of data rates, channel counts, and line spacings [5-10,14-19].

A particularly promising option to generate frequency combs in chip-scale devices relies on Kerr comb generation in micro-ring resonators [27]. In Fig. 3(a), a high-$Q$ resonator is resonantly pumped by a CW laser, thus producing a strong intra-cavity power. Under appropriate pumping conditions, degenerate and non-degenerate four-wave mixing leads to formation of new spectral lines by converting pairs of pump photons into pairs of photons that are up- and downshifted in frequency. Of particular interest for WDM applications are so-called single dissipative Kerr soliton states, which consist of only one ultra-short pulse circulating around the microresonator [28,29]. This leads to broadband frequency combs with particularly smooth spectral envelopes, providing tens or even hundreds of high-quality tones from a single device [6]. An example of a soliton comb spectrum is shown in Fig. 3(b). The soliton comb offers around 110 carriers spaced by approximately 100 GHz within the telecommunication C and L-band (1530 nm … 1610 nm). The line powers range from –11 dBm near the center to – 20 dBm at the edges of the telecommunication window, and the OCNR values range from 48 dB to 40 dB. These values would safely permit link-limited transmission in realistic WDM systems, see Fig. 2.

Using Eq. (8) and the model parameters from Table 1, we investigate the evolution of the OSNR per channel when considering data transmission using the carriers within the C and L band of the soliton frequency comb shown in Fig. 3(b). In addition, we analyze the impact on the OSNR of equalizing the power per line of the frequency comb, see Fig. 3(c)-3(e). If no power equalization is performed to the comb lines, the OSNR variations from channel to channel will be determined by the difference in $P_\ell$ at the output of the frequency comb. Thus, strong channel to channel variations of the OSNR will be observed even after long transmission distances, see Fig. 3(c) "No equalizer". Alternatively, equalization of the different comb lines can be achieved by, e.g., gain equalizing filters in the first amplification stage, see Fig. 1. In this case, after few spans, the ASE noise from the inline amplifiers will bring the noise power of each channel, and thus the OSNR, to a similar value, see Fig. 3(d) "Gain equalizer". However, in practical applications, channel equalization can be easily achieved in the WDM modulator stage, see Fig. 1, by, e.g., the use of optical attenuators or adjusting the driving voltages of the different modulators. In this case, the OSNR for short links is lower compared with the case of no equalization or equalization at the first amplifier stage, see Fig. 3(e) "Power equalizer", due to the extra attenuation of the comb lines after each modulator. However, for long transmission distances, the OSNR values converge to those from Fig. 3(c) and 3(d). Thus, not equalizing the power of the comb lines leads to higher OSNR values for channels at the center of the C and L-band, and lower OSNR values for the channels at the edges of the telecommunication windows, with respect to equalizing the power per channel. Next, we compare the impact of equalling in terms of maximun achievable capacity [30],

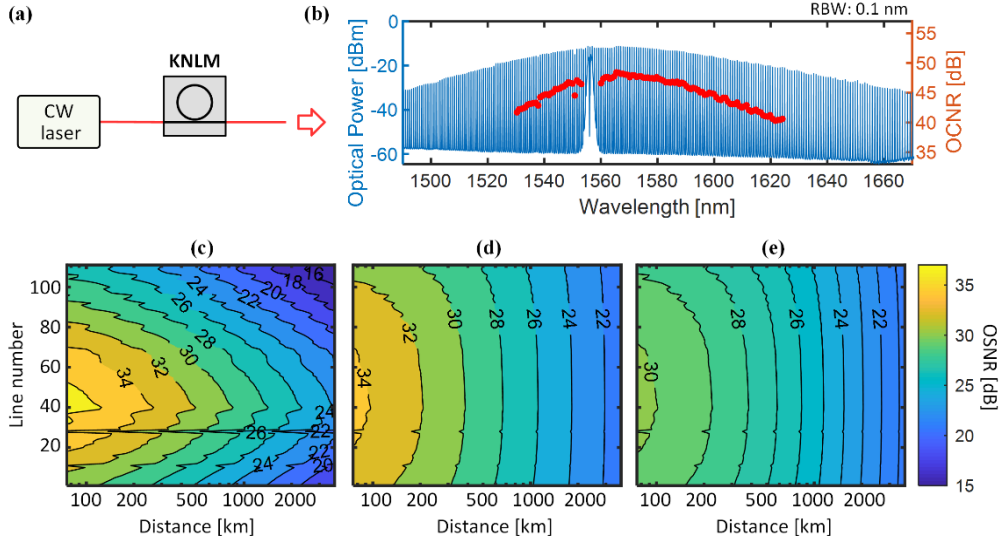

**Fig. 3.** Soliton Kerr combs for massively parallel WDM communications. **(a)** A high-Q Kerr nonlinear microresonator (KNLM) is pumped by a CW laser. Under appropriate conditions, degenerate and non-degenerate four-wave mixing leads to formation of new spectral lines. Of particular interest for optical communications are so-called single dissipative Kerr soliton (DKS) states, for which the superposition of the phase-locked optical tones forms an ultra-short soliton pulse circulating in the cavity. This leads to a comb spectrum with a broadband smooth envelope. **(b)** Spectrum of a single Kerr soliton frequency comb generated in a $Si_3N_4$ microresonator [6]. The comb offers more than 100 carriers in the telecommunication C and L-band (1530 nm … 1610 nm ), spaced by approximately 100 GHz. In the center of the spectrum, the comb line power amounts to − 11 dBm, and the OCNR is approximately 48 dB. **(c)** OSNR as a function of the link distance and the comb line number without equalizing the comb line power; **(d)** equalizing the power per line at the first amplifier stage of Fig. 1; **(e)** equalizing the power per line at the WDM mod. unit from Fig. 1.

$$C = f_r \sum_{\ell=1}^{L} \log_2\left(1 + \text{SNR}_\ell\right) \quad (9)$$

where $f_r$ is the comb line spacing, $L$ is the number of comb lines, and $\text{SNR}_\ell$ is obtained from Eq. (1) with $B = f_r$. Figure 4 shows the capacity as a function of the link distance. The solid lines depict the case where the capacity is maximized for each channel according to its $\text{SNR}_\ell$. This can be done by, e.g., adapting the modulation format for each channel. The dashed lines consider the situation where the channel with the lowest $\text{SNR}_\ell$, $\text{SNR}_{\min}$, dictates the modulation format for all of the channels, and the total capacity is given by

$$C = L f_r \log_2\left(1 + \text{SNR}_{\min}\right) \quad (10)$$

Thus, if the capacity is individually optimized for each channel, the total capacity, $C$, converges to the same value whether or not the comb line powers are equalized. However, when $\text{SNR}_{\min}$ is considered for all channels, the total capacity when flattening the frequency comb is higher than no equalizing for both metro and long-haul links.

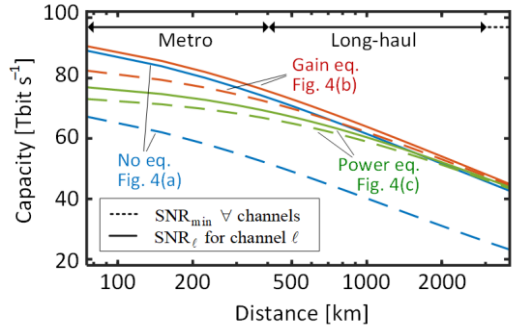

**Fig. 4.** Total capacity of a WDM transmission using Kerr soliton frequency comb. Blue trace: no equalization done. Red trace: the frequency comb is flattened by a gain equalizer in the first amplifying stage. Green trace: the frequency comb is smoothened by optical attenuators in the WDM modulator, Fig. 1. Solid line: the capacity is optimized individually for each channel. Dashed line: The SNR minimum, $\text{SNR}_{\min}$, is used for each channel.

## 4. Summary

We have investigated the influence of the comb line power and optical carrier-to-noise power ratio (OCNR) on the OSNR for comb-based WDM transmission systems. We distinguish two regimes of operation depending on whether the comb source or the transmission link limits the performance of the system. Finally, we have analyzed the achievable OSNR and channel capacity when using the tones of a soliton frequency comb as WDM carriers.

**Acknowledgements**

This work was supported by the EU-FP7 project BigPipes (# 619591), by the European Research Council (ERC Consolidator Grant 'TeraSHAPE', # 773248), by the Deutsche Forschungsgemeinschaft (DFG) through the Collaborative Research Center 'Wave Phenomena: Analysis and Numerics' (CRC 1173, project B3 'Frequency combs'), by the Karlsruhe School of Optics & Photonics (KSOP), by the Helmholtz International Research School for Teratronics (HIRST), and by the Alfried Krupp von Bohlen und Halbach Foundation. P.M.-P. was supported by the Erasmus Mundus doctorate programme Europhotonics (grant number 159224-1-2009-1-FR-ERA MUNDUS-EMJD).